\documentclass[12pt,preprint]{aastex}
\usepackage{url}
\usepackage{natbib}
\usepackage{aas_macros}
\usepackage{amsmath}
\shorttitle{Dynamics of Turbulent Convection and Convective Overshoot in a Moderate Mass Star}

\title{Dynamics of Turbulent Convection and Convective Overshoot in a Moderate Mass Star}

\author{I.~N. Kitiashvili$^{1}$, A.~G. Kosovichev$^{2}$, N.~N. Mansour$^{1}$, A.~A. Wray$^{1}$}
\affil{$^1$NASA Ames Research Center, Moffett Field, Mountain View, CA 94035, USA}
\altaffiltext{1}{e-mail: irina.n.kitiashvili@nasa.gov}
\affil{$^2$New Jersey Institute of Technology, Newark, NJ 07102, USA}

\begin{document}

\begin{abstract}
Continued progress in observational stellar astrophysics requires a deep understanding of the underlying convection dynamics. We present results of realistic 3D radiative hydrodynamic simulations of the outer layers of a moderate mass star (1.47~M$_\odot$), including the full convection zone, the overshoot region, and the top layers of the radiative zone. The simulation results show that the surface granulation has a broad range of scales, from 2 to 12~Mm, and that large granules are organized in well-defined clusters, consisting of several granules.  Comparison of the mean structure profiles from 3D simulations with the corresponding 1D standard stellar model shows an increase of the stellar radius by  $\sim 800$~km, as well as significant changes in the thermodynamic structure and  turbulent properties of the ionization zones. Convective downdrafts in the intergranular lanes between granulation clusters reach speeds of more than 20~km s$^{-1}$, penetrate through the whole convection zone, hit the radiative zone, and form a 8~Mm thick overshoot layer.
Contrary to semi-empirical overshooting models, our results show that the 3D dynamic overshoot region consists of two layers: a nearly adiabatic extension of the convection zone and a deeper layer of enhanced subadiabatic stratification. This layer is formed because of heating caused by the braking of the overshooting convective plumes. This effect has to be taken into account in stellar modeling and the interpretation of asteroseismology data. In particular, we demonstrate that the 3D model can qualitatively explain deviations from the standard solar model found by helioseismology.

\end{abstract}

\section{Introduction}
Until recently only 1D stellar models based on the mixing-length theory (MLT) were available. However, rapidly growing computational capabilities have enabled 3D stellar dynamics simulations on both global \citep[e.g][]{Meakin2007,Charbonneau2013,Guerrero2013,Featherstone2015} and local scales \citep[e.g.][]{Brummell2002,Trampedach2010,Beeck2012,Kitiashvili2012,Magic2013,Trampedach2013}. 

The overshoot region at the bottom of the convection zone plays a particularly important role in stellar dynamics and magnetism. Helioseismology has found that this region coincides with the tachocline, a zone of  strong rotational shear \citep{Kosovichev1996,Elliott1999} that is considered to be a primary source of the solar dynamo. Our basic knowledge about solar tachocline properties comes from helioseismic inversions of p-mode frequencies of medium angular degree \citep[e.g.][]{Kosovichev1997}, which find a pronounced peak in the sound-speed deviations from the standard evolutionary model at the bottom of the convection zone \citep{Christensen-Dalsgaard1996} as well as deviations in other properties, such as density, entropy gradient, and the adiabatic exponent \citep{Kosovichev1999}. Attempts to explain these deviations in terms of rotationally induced mixing, variations of opacity, and 1D `semi-empirical' overshooting models have not been successful \citep[for a recent results and references, see the paper of][]{Christensen-Dalsgaard2011}. Realistic 3D simulations, which could shed light on tachocline structure and properties, currently cannot be performed because of the large depth of the solar convection zone and long relaxation time. However, as suggested by \citet{Freytag1996}, such simulations can be performed for more massive F- and A-type stars with much shallower  outer convection zones.  

We present new results of local 3D (non-MLT) radiative hydrodynamic simulations for a star with mass 1.47~M$_{\odot}$, including the whole convection zone, the overshoot region, and the outer part of the radiative zone. We discuss the physical properties of the turbulent convection and the structure and dynamics of the overshoot region. Comparison with the corresponding 1D standard stellar model reveals a striking similarity to the deviations of the solar stratification from the standard solar model previously found using helioseismology. While this comparison is only qualitative, it reveals important features of convective overshooting that were missing in previous studies. 

\section{Computational setup}
The stellar convection simulations are performed using the 3D radiative MHD code `StellarBox' \citep{Wray2015} that was specially developed for realistic modeling of solar and stellar convection. The simulations are Large-Eddy Simulations (LES) that use the Smagorinsky model \citep{Smagorinsky1963,Moin1991} to describe subgrid-scale turbulence (which also supplies diffusion necessary to stabilize the numerical scheme for  solving the PDEs). Radiative transfer between fluid elements is calculated using a 3D, multi-spectral-bin, long-characteristics method, in which the ray-tracing transport is calculated with the \citet{Feautrier1964} method  for  14 rays. The OPAL opacity wavelength-dependent code and equation-of-state tables  were used assuming a solar-type composition:  $X=0.702, Y=0.28,  Z=0.018$ \citep[see][for details]{Iglesias1996,Rogers1996}.

The initial conditions are calculated using  the stellar structure model obtained with the 1D stellar evolution code CESAM \citep{Morel1997,Morel2008}  using the same composition and OPAL tables, and a standard value of the mixing-length parameter  $\alpha=$1.8. The turbulent pressure and overshoot parameters are set zero.
With increasing stellar mass the thickness of the convection zone decreases. This allows us to investigate the dynamics of the entire convection zone of a moderate mass star. For this study, we selected a star with the following properties: $\rm M=1.47~M_\odot$, $\rm T_{eff}=7063$~K, $g=10^{4.279}$~cm  s$^{-2}$, ${\rm L/L}_{\odot}=4.731$, $\rm R=1.456~R_{\odot}$. The depth of the convection zone is: $z_{\rm cz}=-28.5$ Mm.
The 1D stellar model was extended by adding an isothermal 1-Mm atmospheric layer. This allowed us to accurately model the radiative cooling of the surface layers by keeping the optical depth at the upper boundary at about $10^{-2}$, despite the expansion of the upper convection zone caused by the turbulent dynamics.

The simulation domain used in the current simulations is $102.4\times 102.4 \times 51$~Mm in size ($1024\times 1024\times 512$ grid points). The computational domain covers 1~Mm of the  atmosphere and 50.5~Mm of the stellar interior. This includes the entire convection zone and the upper part of the radiation zone. The grid spacing is 100~km in the horizontal directions and variable in the vertical direction, from  28.6~km at the photosphere and above to 183~km at the bottom boundary of the domain. The boundary conditions are periodic in the horizontal directions. The bottom boundary is closed for flows and open for the radiative energy flux. To suppress vorticity generation near the bottom boundary we increased the dissipation by imposing a viscous layer (occupying 4 vertical grid points) near the bottom boundary. The upper boundary condition is formulated in terms of the Riemann characteristics and is open for mass, momentum, and energy fluxes and also for the radiation flux. For other computational details and performance of the `SolarBox'  see \citet{Wray2015}. 

\begin{figure}
\begin{center}
\includegraphics[scale=0.75]{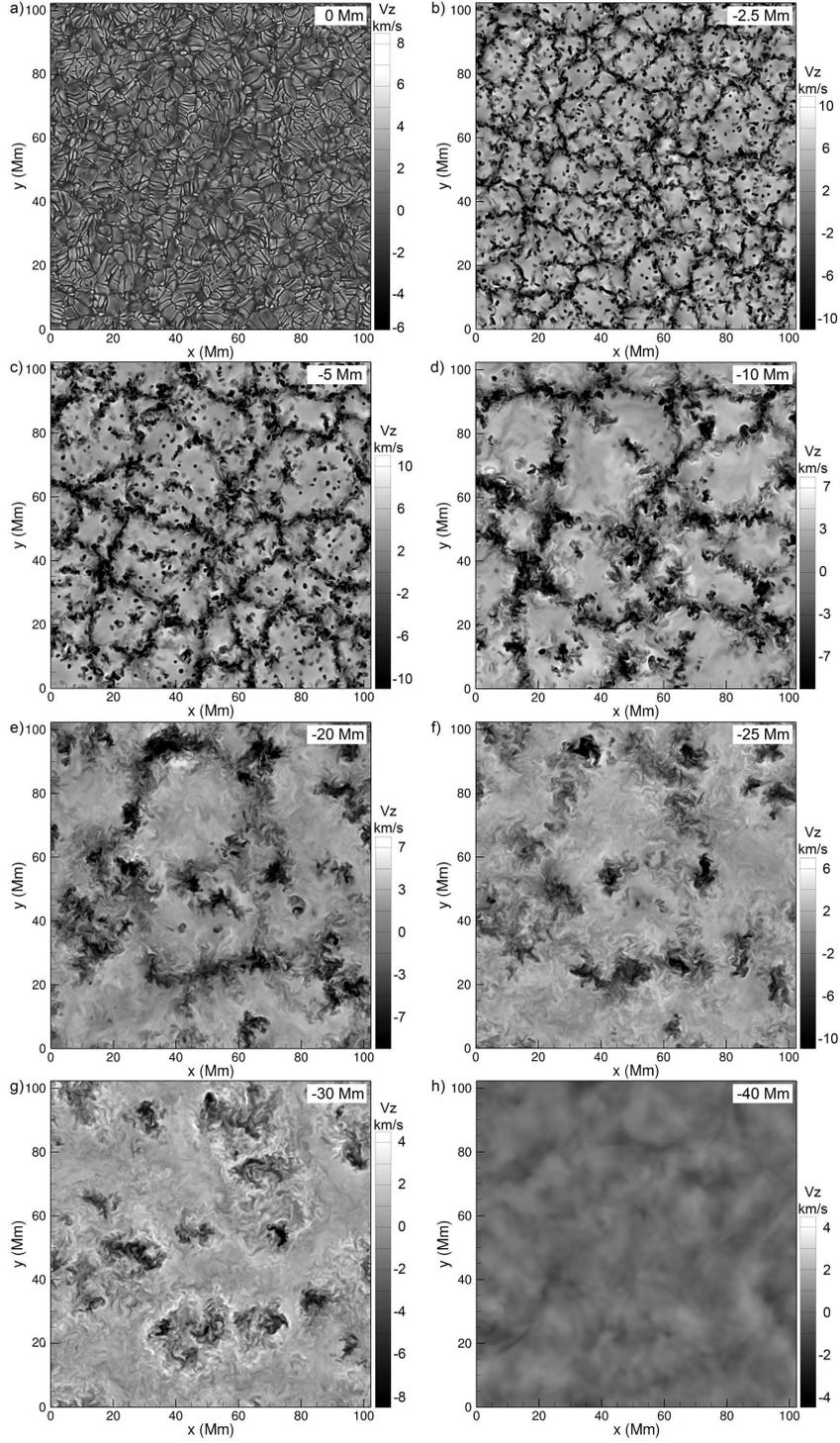}
\end{center}
\caption{Vertical velocity distribution for the simulated star (1.47~M$_{\odot}$) at eight depths throughout the convective zone and into the radiative zone, $0, 2.5, 5, 10, 20, 25, 30$, and $40$~Mm below the  surface ($r=R$). To display small-scale details the color scales are chosen shorter than the full velocity range. \label{fig:XY} }
\end{figure}

\begin{figure}[t]
\begin{center}
\includegraphics[scale=1]{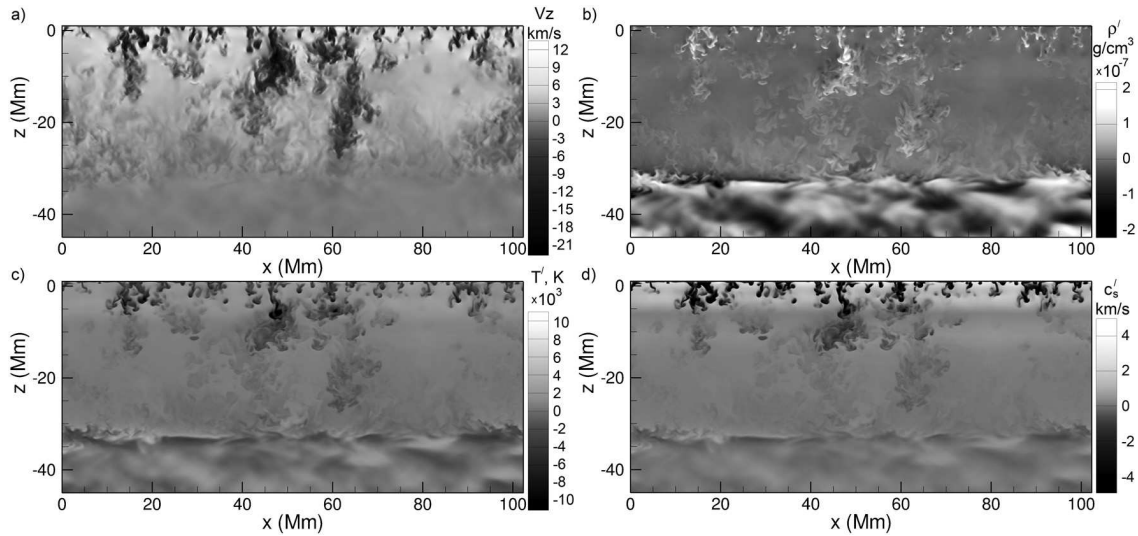}
\end{center}
\caption{Vertical slices of: $a$) the vertical velocity, $V_z$; and fluctuations (relative to the mean 3D profiles of Fig.~\ref{fig:model_ML}) of $b$) density, $\rho'$; $c$) temperature, $T'$; and $d$) sound speed, $c_s'$, for the modeled star (1.47~M$_{\odot}$). Depth $z$ is defined as: $z=r-R$.
\label{fig:XZ} }
\end{figure}

\begin{figure}[!h]
\begin{center}
\includegraphics[scale=0.64]{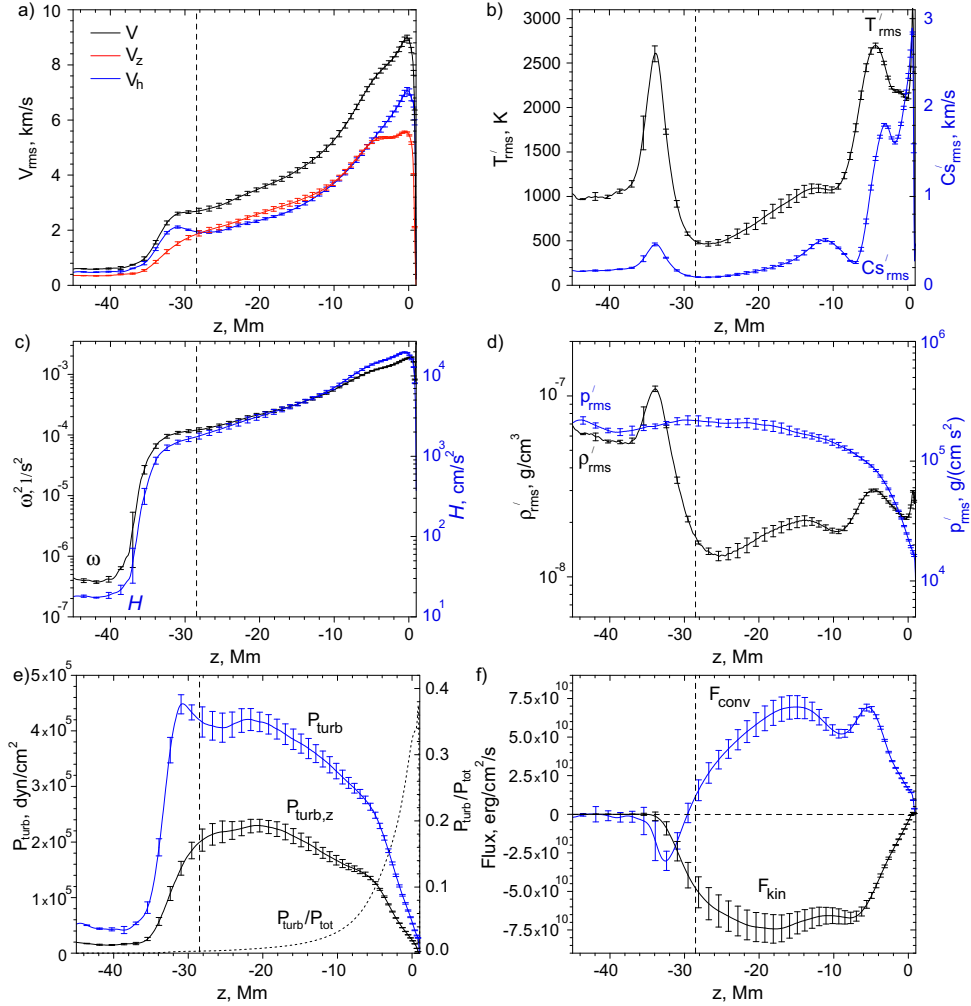}
\end{center}
\caption{Vertical profiles, obtained from the 3D numerical simulation of a 1.47\,M$_\odot$ F-type star: $a$) {\it rms} of velocity $V$ (black), vertical $V_z$ (red) and horizontal $V_h$ (blue) components of velocity; $b$) {\it rms}  of temperature $T^\prime$ (black) and  sound speed $c_S^\prime$ (blue)  perturbations;  $c$) enstrophy $\omega$ (black) and helicity $H$ (blue); $d$) {\it rms}  of density $\rho^\prime$ (black) and gas pressure $P^\prime$ (blue) perturbations;  $e$) turbulent $P_{\rm turb}$  (blue curve), turbulent pressure of the vertical motions $P_{\rm turb}$  (black), and the ratio of turbulent pressure $P_{\rm turb,z}$ to total pressure $P_{\rm turb}+P$;  and  $f$) convective energy flux $F_{\rm conv}$ (blue curve) and kinetic energy flux $F_{\rm kin}$ (black), calculated  according to the formulation of \citet{Nordlund2001}. Vertical dashed lines indicate the bottom boundary of the convection zone of the corresponding 1D stellar model: $z_{\rm cz}=-28.5$ Mm.
\label{fig:XZ1} }
\end{figure}

\section{Turbulent dynamics of the stellar convection}
The stellar photosphere is covered by cells, that, like granules on the Sun, are characterized by upflows in the middle and narrow downflows in the surrounding lanes (Fig.~\ref{fig:XY}$a$, and online movie). 
Large-scale fluctuations in this movie reflect propagation of acoustic waves excited by the turbulent convection \citep[e.g.,][]{Stein2001,Jacoutot2008a,Kitiashvili2011,Kitiashvili2011a}. 
The size of stellar granules varies from $\sim 2 $~Mm to $\sim 12$~Mm, and clusters of several granules (resembling mesogranulation) are evident. Most granules of an intermediate size are stretched by larger-scale upflows. The multi-scale property of the convection vanishes in subsurface layers, where the scale of the convective eddies increases with depth (Fig.~\ref{fig:XY}$a-f$) and reaches a size of 40--45~Mm at about 25~Mm depth (near the bottom of the convection zone).  Deeper down the influence of the overshoot region becomes more and more noticeable as prominent fine structures of upflows are formed around edges of strong downflows. 
Figure~\ref{fig:XY}$g$ illustrates a horizontal slice of the overshoot region at 30~Mm depth, where high-speed downflows (Fig.~\ref{fig:XZ}$a$) penetrate through the whole convection zone and hit the radiative zone, causing splashes in the horizontal directions and generating numerous small-scale turbulent helical structures and upflows around the plumes.
 We define the overshoot region as a zone of penetration of convective flows beneath the adiabatic boundary at $z_{\rm cz}$ (see Sec.~4 and 5 for more detail). Below the overshoot region, at the depth of 40~Mm (Fig.~\ref{fig:XY}$h$), in the radiative zone (which is a convectively stable layer), the velocity perturbations quickly decrease, and the dynamics is dominated by internal gravity waves.

Our simulation results show that narrow high-speed downdrafts (Fig.~\ref{fig:XZ}$a$) driven by surface radiative cooling can often penetrate through the whole convection zone, but they quickly stop in the overshoot layer. The downflows are accelerated up to 21~km s$^{-1}$ and initiate strong perturbations in the radiative zone. Such perturbations are clearly visible in the density, temperature, and sound speed fluctuations (Fig.~\ref{fig:XZ}$b-c$). The downdrafts are associated with positive density perturbations and negative temperature perturbations. However, the penetrating downflows locally heat the overshoot region. The downdraft impacts also lead to the excitation of internal gravity waves ($g$-modes) in the radiative interior. 

To summarize the dynamical structuring from the stellar photosphere down to the radiative zone, it is convenient to consider the mean vertical profiles of the fluctuations for various turbulence quantities, and the convective and kinetic energy flux (Fig.~\ref{fig:XZ1}). The {\it rms} velocity profiles show the relative strength of the vertical and horizontal flows. The horizontal flows gradually decrease in strength from the photosphere to the bottom of the convection zone. In the overshoot region, the amplitude of the horizontal velocities increases slightly due to splashing of downdrafts. 

The {\it rms} vertical velocity profile shows a strong deviation from the horizontal flows near the stellar surface layer. In particular, an additional broad bump at about 5~Mm below the surface corresponds to one of the characteristic scales of the granulation layer.  In deeper layers, the vertical velocity gradually decreases. It is interesting to note that, in the range of depths of 6 -- 10~Mm, the horizontal and vertical flows behave very similarly.  Below 10~Mm the fluctuations of the vertical velocities are stronger than the horizontal ones, probably reflecting the influence of the penetrating high-speed downdrafts. In the overshoot layer, the vertical velocity sharply decreases. 

The distribution of temperature fluctuations with depth (Fig.~\ref{fig:XZ1}$b$, black curve) shows the strongest variations near the surface layers and in the overshoot region. Despite the similarity of the fluctuation amplitudes, their nature is different. The near-surface fluctuations (peaked at $\sim 5$~Mm) are mostly related to strong radiative cooling in the intergranular lanes and are associated with downflows that determine the granulation scales. A sharp increase in temperature fluctuations between $-10$ and $-5$~Mm corresponds to the HeII ionization zone. In this region the {\it rms}  of vertical velocity (Fig.~\ref{fig:XZ1}$a$) and the convection energy flux (Fig.~\ref{fig:XZ1}$f$) also increase, indicating an enhancement of turbulent convection in the ionization zone, perhaps, confirming theoretical predictions of \citet{Rast1993}. 

The strong fluctuations in temperature at $z\approx -34$~Mm, which are associated with the overshoot region, reflect the strong local heating in these layers due to plasma compression by downdrafts that penetrate through the convection zone.  The {\it rms} sound speed perturbations  (Fig.~\ref{fig:XZ1}$b$, blue curve) have similar shape, but their amplitude in the overshoot region is much smaller than in the subsurface layers.
The mean profiles of the enstrophy and kinetic helicity density (Fig.~\ref{fig:XZ1}$c$) gradually decrease with depth through the convection zone, and then sharply decrease at the bottom of the overshoot region. These quantities are important for the dynamo action.

The vertical profiles for the {\it rms} fluctuations of density and gas pressure (Fig.~\ref{fig:XZ1}$d$) are qualitatively different. The density fluctuations show variations that are qualitatively similar to the temperature fluctuations, whereas the gas pressure fluctuations gradually increase from the surface to about 20~Mm below the surface, and then remain almost constant in the deeper layers. However, the total turbulent pressure, $P_{\rm turb} $, displays a prominent peak in the upper half of the overshoot region (Fig.~\ref{fig:XZ1}$e$) while the turbulent pressure of vertical flows, $P_{\rm turb,z}$, gradually decreases. The convective enthalpy flux, $F_{\rm conv}$, calculated following the formulation of \citet{Nordlund2001} (Eqs 22--23), is negative in the overshoot region as suggested by non-local mixing-length theories \citep[e.g.][]{Shaviv1973}.

\begin{figure}
\begin{center}
\includegraphics[scale=0.8]{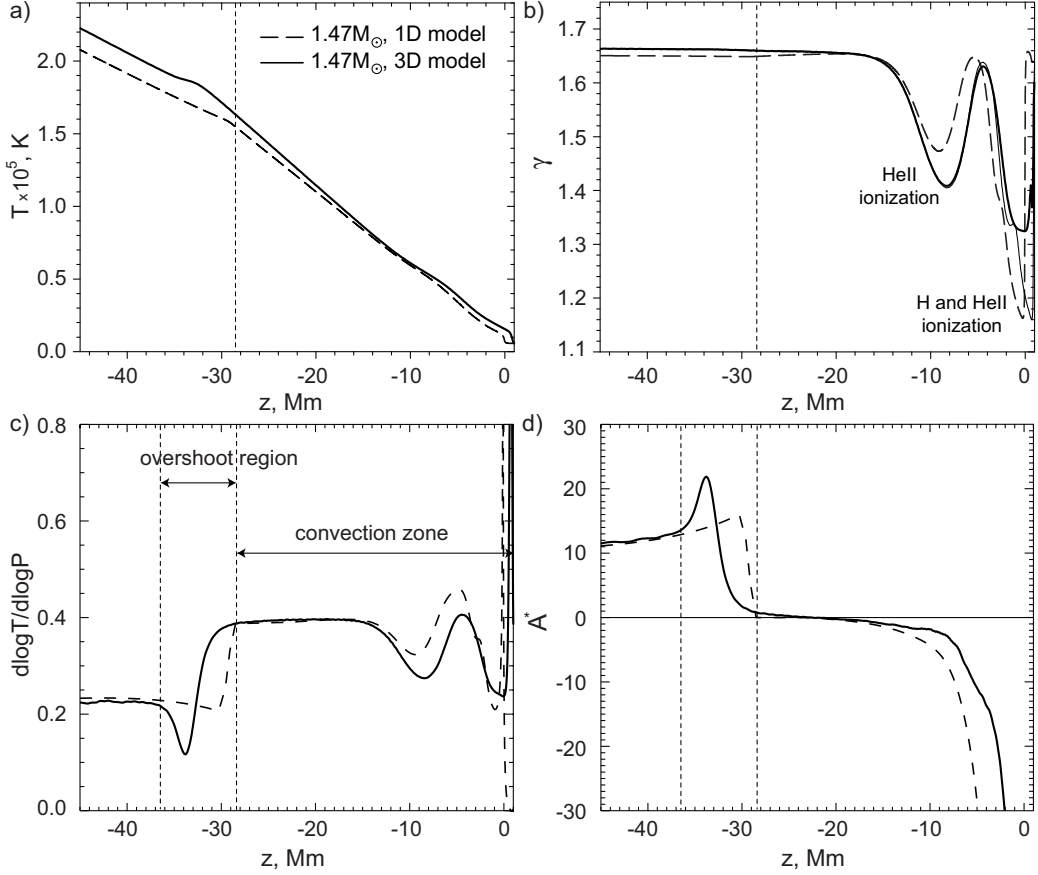}
\end{center}
\caption{Comparison of the 1D interior structure of a moderate mass star (M=1.47M$_\odot$) calculated from mixing-length theory (dashed curves) and from the 3D simulation (solid curves): $a$) the temperature, $T$; $b$) the adiabatic exponent, $\gamma$; $c$) the temperature gradient, $\nabla=\dfrac{d\log T}{d\log P}$;  $d$) the Ledoux parameter of convective stability, $A^*=\dfrac{1}{\gamma}\dfrac{d\log P}{d\log r}-\dfrac{d\log\rho}{d\log r}$. Vertical dotted lines indicate the bottom of the convection zone and the extent of the overshoot region in the 3D simulation. The mean profiles of the 3D simulation are calculated by averaging in the horizontal directions and over a 1 hour interval. The thin curve in panel $b$) shows the value of $\gamma$ calculated from the mean density and internal energy profiles of the 3D model.} \label{fig:model_ML}
\end{figure}

\section{Comparison of the 1D model and 3D simulations}

In 1D stellar evolution models the convection zone structure is calculated using mixing-length theory \citep[MLT,][]{Bohm-Vitense1958}, which describes convection as buoyant bubbles traveling in mixing length steps from deep convective layers  to the surface. However, numerical simulations have shown that a significant role in solar and stellar convection is played by high-speed downdrafts driven by surface cooling. Therefore, it is not surprising that there are significant differences between the MLT-based models and the corresponding mean properties of the 3D simulations of turbulent convection. 

In the case of a star with mass 1.47~M$_\odot$, we can investigate the differences between the 1D MLT model and 3D LES simulations through the whole convection zone. Figure~\ref{fig:model_ML} shows the mean profiles of temperature $T$, adiabatic exponent $\gamma=(\partial\log P/\partial\log\rho)_{\rm ad}$,  temperature gradient $\nabla=d\log T/d\log P$, and the Ledoux parameter of convective stability, $A^*=1/\gamma(d\log P/d\log r) - d\log\rho/d\log r$, where $P$ is the gas pressure, $\rho$ the mass density, and $r$ the radius, for the 1D model and 3D simulations.

In the 3D stellar simulations the convection zone expands in both directions: about 800~km up into the atmosphere with a corresponding increase of the stellar photospheric radius. Such increase of the stellar radius due to turbulent dynamics may be relevant to resolving the discrepancy between the solar seismic and photospheric radii  \citep{Schou1997}.
In the 3D simulations, the mean profile of the temperature is slightly higher than in the 1D model, but the deviation is strongest in the near-surface layers where the effects of radiative losses and turbulent dynamics are strong. At about 10~Mm, where the HeII ionization zone starts, the temperature discrepancies are decreasing and become smallest at the lower boundary of the zone. In the deeper layers the difference in the temperature profiles increases again and is highest in the overshoot region. 
Our simulation results show significant changes in the value of $\gamma$ for the hydrogen and helium ionization zones (Fig.~\ref{fig:model_ML}$b$): decreasing for the hydrogen and increasing for the HeII ionization zones. For comparison we plot the values of $\gamma$ (thin curve) calculated from the horizontally averaged values of the density and internal energy of the 3D model. The differences of the curve from the 3D model (thick curve) in the near-surface layers indicate that the nonlinear dynamical fluctuations caused by turbulence and waves play an important role in the thermodynamic properties.
These deviations from the 1D models require further investigation, and have to be taken into account when estimating helium abundances by helio- and astero-seismology.

\begin{figure}[t]
\begin{center}
\includegraphics[width=\linewidth]{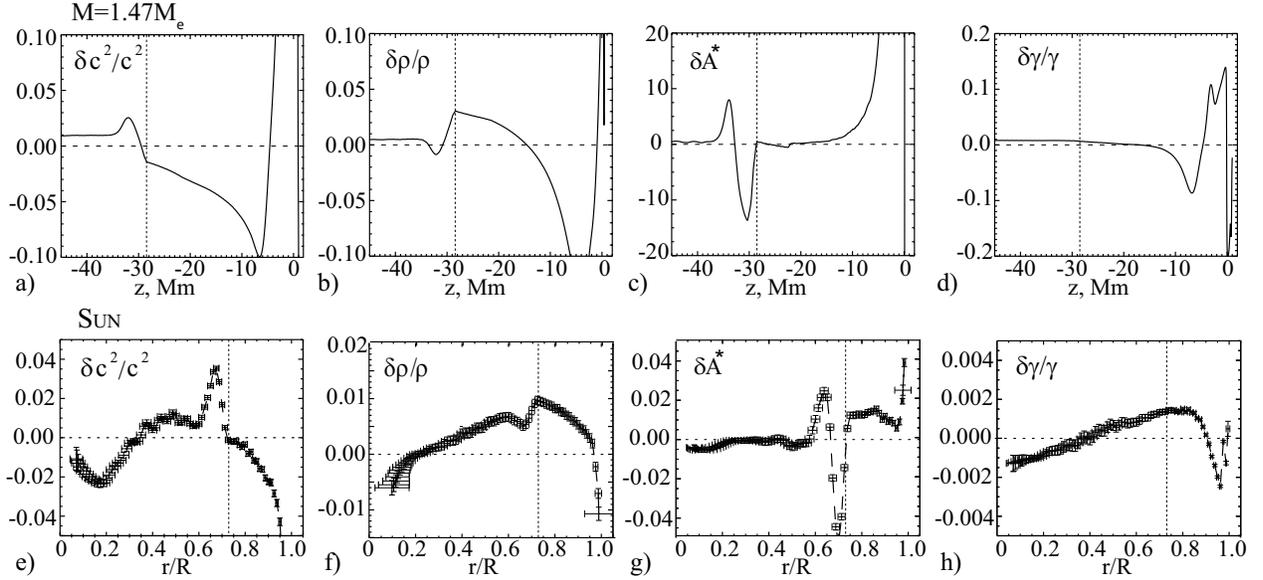}
\end{center}
\caption{The deviations between the 3D simulation and 1D model of a star with mass $M=1.47~M_\odot$  as a function of depth, $z=r-R$,  for:  $a$) the squared sound speed, $\delta c^2/c^2$; $b$) density, $\delta\rho/\rho$; $c$) the Ledoux parameter of convective stability, $A^*$; and $d$) the adiabatic exponent, $\gamma$. Panels $e-h$) show the corresponding deviations of the solar properties obtained by helioseismology inversion \citep{Kosovichev1999,Kosovichev2011} from the  1D standard solar model  \citep{Christensen-Dalsgaard1996}. Vertical dotted lines show the location of the bottom boundary of the convection zone.}\label{fig:Cs} 
\end{figure}
 
\section{Structure of the overshooting region}

The region of convective overshoot is defined in terms of the temperature gradient as a transition zone between the adiabatic and radiative gradients \citep[e.g.][]{Christensen-Dalsgaard2011}. It is indicated by two vertical lines in Fig.~\ref{fig:model_ML}$c-d$,  from $-28.5$~Mm to $-36.5$~Mm in depth,  and is approximately 1/2 of the pressure scale height. This closely corresponds to the region of negative enthalpy flux (Fig.~\ref{fig:XZ1}$h$), which is a distinct feature of convective overshooting. The  simulation results clearly show that the overshooting region consists of two parts: the upper part represents an almost adiabatic extension of the convection zone, while the lower part shows a significant peak of increased subadiabaticity. A similar two-layer structure was obtained in 2D anelastic simulations of solar convection by \citet{Rogers2005}. 
However, most previous `semi-empirical' models \citep[see][for a review]{Christensen-Dalsgaard2011} described the overshoot region as an extension of the depth of the convection zone with a smooth monotonic transition between the adiabatic and radiative temperature (or entropy) gradients. The numerical simulations presented in the previous section show that the braking of penetrating convective plumes results in heating of the overshoot region. This leads to a sharp decrease of the temperature gradient in the lower part of the overshoot region. This effect has to be taken into account in the construction of 1D models of convective overshoot based on modification of the temperature gradient in the standard solar and stellar models.

\section{Relative structure variations}
The sound-speed variations are of particular interest because asteroseismic estimations based on $p$-mode frequencies are mostly sensitive to the sound-speed profile. Usually, the sound-speed profile calculated from a 1D stellar model is used as a reference; therefore it is important to investigate the deviations in sound speed between the 3D simulation and 1D model. Figure~\ref{fig:Cs}$a$ shows the relative sound-speed difference.
This profile reveals strong deviations from the 1D model. These deviations are qualitatively similar to the findings of global helioseismology analysis reproduced in Fig.~\ref{fig:Cs}$e$ \citep[e.g.][]{Kosovichev1997}. In particular, the 3D simulations reproduce the characteristic `bump' at the bottom of the convection zone ($z \approx -33$~Mm in Figure~\ref{fig:Cs}$a$). This bump corresponds to the overshoot layer, which is about 8~Mm thick. Our simulation results show a qualitative agreement for this bump's shape with helioseismology results, but the amplitude of the variations in the stellar simulations is much larger than in the solar measurements,  obviously, because the convection is much more vigorous  in the F-type star than in the Sun.
Nevertheless, the surprising similarity supports an idea about the influence of the convective overshoot in the solar sound-speed inversions. In other panels of Fig.~\ref{fig:Cs} we illustrate the qualitative agreement for the relative variations of density, the parameter of convective stability, and for the adiabatic exponent. The good correspondence leads us to the suggestion that the structure of the overshooting region in the Sun is similar to the overshoot structure obtained in our simulations.

\section{Conclusion}
We have performed, for the first time, detailed 3D fully compressible radiative  hydrodynamic simulations of the convection zone dynamics of a moderate mass  F-type star through its whole depth and into the transition layer between the convection and radiative zones (`overshoot zone'). A characteristic feature of the convection zone is a co-existence of several  granulation scales at the photosphere. The convection pattern represents well-defined clusters of several granules. 
In deeper layers of the convection zone, the scale of the convection patterns gradually increases (Fig.~\ref{fig:XY}). The downflows in the intergranular lanes of the bigger granules penetrate through the whole convection zone, reaching velocities of more than 20~km s$^{-1}$, close to the local speed of sound (Fig.~\ref{fig:XZ}). These downdrafts penetrate into the radiative zone, form an overshoot layer, and cause local heating, thus increasing the sound speed and initiating strong density variations that can be a source of internal gravity oscillations ($g$-modes).  

Comparison of the initial 1D model calculated from the stellar evolution code (CESAM) with the corresponding mean profiles from the 3D numerical simulation shows that in the 3D simulations the stellar radius is increased by about 800~km, and the convection zone is expanded in depth due to the development of an overshoot region about 8~Mm (1/2 of the pressure scale height) thick (Fig.~\ref{fig:model_ML}).  The overshoot region is characterized by a negative enthalpy flux and, most importantly, has a two-layer structure:   the upper part represents an almost adiabatic extension of the convection zone, while the lower part shows a significant peak of increased subadiabaticity, caused by heating of the overshooting region due to  braking of penetrating convective plumes. In addition, the simulations reveal a layer of enhanced turbulence, associated with the ionization zones, which affects variations of the adiabatic exponent. 

The deviations of the mean properties between the 3D and 1D models are remarkably similar to the corresponding deviations between helioseismology inversions and the standard solar model  (Fig.~\ref{fig:Cs}), indicating that, perhaps, the helioseismology results including the characteristic sound-speed bump may be explained by dynamical effects of solar convection and overshooting.

 
\end{document}